\documentclass[12pt,showpacs,a4paper,english]{revtex4}
\usepackage[T1]{fontenc}
\usepackage[latin1]{inputenc}
\usepackage{babel}
\usepackage{graphics}
\usepackage{setspace}

\begin{document}

\title{Scalar susceptibility and chiral symmetry restoration in nuclei.}

\author{G. Chanfray\protect\( ^{1}\protect  \), M. Ericson\protect\( ^{1,2}\protect  \), P.A.M. Guichon\protect\( ^{3}\protect  \)}

\affiliation{{ \protect\( ^{1}\protect \)IPNLyon, IN2P3-CNRS et UCB
 Lyon I, F69622 Villeurbanne Cedex} }

\affiliation{{ \protect\( ^{2}\protect \)Theory division, CERN, CH12111 Geneva} }

\affiliation{{ \protect\( ^{3}\protect \)SPhN/DAPNIA, CEA-Saclay, F91191 Gif
sur Yvette Cedex} }

\begin{abstract}
We study the nuclear modification of the scalar QCD susceptibility,
calculated as the derivative of the quark condensate with respect
to the quark mass. We show that it has two origins. One is the low
lying nuclear excitations. At normal nuclear density this part is
constrained by the nuclear incompressibility. The other part arises
from the individual nucleon response and it is dominated by the pion
cloud contribution. Numerically the first contribution dominates.
The resulting increase in magnitude of the scalar susceptibility at
normal density is such that it becomes close to the pseudoscalar susceptibility,
while it is quite different in the vacuum. We interpret it as a consequence
of chiral symmetry restoration in nuclei.
\end{abstract}

\maketitle

\section{Introduction}

The chiral phase transition is a hot topic of QCD. The attention has
focused in particular on a baryonic rich environement, the simplest
one being ordinary nuclei. In view of the difficulties of the lattice
methods to treat baryonic matter, studies based on models have revealed
their interest. In the first stage the order parameter, which is the
quark condensate, was investigated\cite{DL90,CFG92}. With the realization
that the amount of restoration is large, as the order parameter has
decreased by about 35\% at ordinary nuclear density, the interest
has also focused on precursor effects linked to the partial restoration
of chiral symmetry in the form of dropping hadron masses\cite{BR91,B96,DEGT00}
or axial-vector correlator mixing\cite{DEI90,CDE98}. The last topic
to attract attention is the question of the susceptibilities in QCD\cite{HKS99,CE02}.
In the broken symmetry phase, where the order parameter introduces
a privileged direction, the susceptibility is splitted as in magnetism
in the parallel susceptibility, along the magnetization axis, and
the perpendicular one. In QCD these are the scalar (\( \chi _{S} \))
and pseudoscalar (\( \chi _{PS} \)) susceptibilities related to the
fluctuations of the scalar and pseudoscalar quark density, respectively.

\begin{figure}
{\centering \resizebox*{0.6\textwidth}{!}{\includegraphics{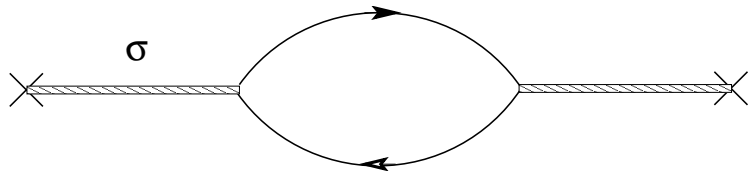}} \par}

\caption{\label{Fig_sig_ph}Modification of the \protect\( \sigma \protect \)
propagator by the particle-hole polarisation propagator. The cross
represents the condensate}
\end{figure}

\begin{figure}
{\centering \resizebox*{0.6\textwidth}{!}{\includegraphics{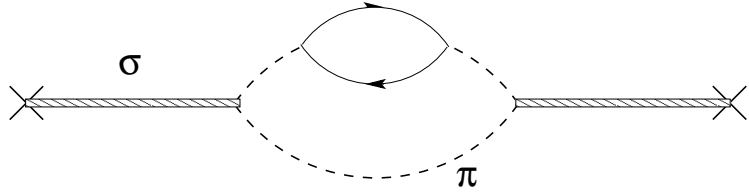}} \par}

\caption{\label{Fig_sig_2pi}Modification of the \protect\( \sigma \protect \)
propagator by the in medium 2\protect\( \pi \protect \) propagator.
The cross represents the condensate}
\end{figure}

The in-medium QCD susceptibilities have been discussed by Chanfray
and Ericson\cite{CE02}. For the scalar one they used the linear sigma
model which provides the coupling of the quark scalar density fluctuations
to the nucleonic ones through sigma exchange. They ignored the coupling
to the pion density fluctuations, expected to have a smaller influence.
Their approach is basically a dispersive one, with the introduction
of the in-medium scalar spectral function. In terms of graphs their
effect corresponds to the one of Figure \ref{Fig_sig_ph} and the
dressing by the pion lines to that of Figure \ref{Fig_sig_2pi}.

The present work uses a totally different approach which relies on
the very definition of the longitudinal susceptibility as the derivative
of the order parameter with respect to the perturbation responsible
for the explicit symmetry breaking. In magnetism it is the external
magnetic field. In QCD it is the quark mass and we have the generic
definition of the scalar susceptibility:\begin{equation}
\label{Eq-1}
\chi _{S}=\frac{\partial }{\partial m_{q}}<\bar{q}q>,
\end{equation}
 where \( q \) is the quark field. The two methods are known to be
equivalent in their principle. However in practice, in the dispersive
approach, truncations are made in the intermediate accessible states.
In fact our new derivation incorporates terms as well as interference
effects which were previously absent. The present study is also free
from the specific features of the linear sigma model.

In Section \ref{section-2}, we show how Eq\.{ }(\ref{Eq-1}) leads
to a natural decomposition of \( \chi _{S} \) into a vacuum contribution
\( \chi _{S}(vac.), \) a contribution noted \( \chi _{S}^{N} \)
whis is related to the \emph{nucleonic} excitations, and a contribution
noted \( \chi _{S}^{nuclear} \) which is related to the \emph{nuclear}
excitations. In Section \ref{section-3} we use the Fermi gaz model
to estimate \( \chi _{S}^{nuclear} \) and we compare the result with
the one obtained previously with the linear sigma model\cite{CE02}.
We conclude that \( \chi _{S}^{nuclear} \) depends essentially on
the zero momentum particle-hole propagator which we relate to the
nuclear incompressibility. In Section 4 we study the nucleonic contribution
\( \chi _{S}^{N} \) and we find that it is dominated by the pion
cloud, the quark core giving only about 6\% of \( \chi _{S}^{N} \).
The model dependance of the pion cloud contribution is moderated by
the fact that, within a factor of order two, its value is fixed by
the leading non analytic piece of the sigma term. \textbf{}In Section
\ref{Numerics} we present some numerical estimates based on the results
of Sections 2, 3 and we discuss their implications concerning the
restoration of Chiral symmetry in nuclei.

\section{Nuclear susceptibility\label{section-2} }

Since our aim is to evaluate the modification of the susceptibility
with respect to its vacuum value we note \( <\bar{q}q(\rho )> \)
the in-medium value of the quark condensate and we introduce:\begin{equation}
\label{Eq.2}
\chi _{S}(\rho )=\frac{\partial }{\partial m_{q}}<\bar{q}q(\rho )>
\end{equation}
 where \( \rho  \) is the nuclear density.

For a collection of independent nucleons, the in-medium condensate
\( <\bar{q}q(\rho )> \) writes:

\begin{equation}
\label{Eq.3}
<\bar{q}q(\rho )>=<\bar{q}q(vac.)>\left[ 1-\frac{\Sigma _{N}\rho _{S}}{f_{\pi }^{2}m_{\pi }^{2}}\right] .
\end{equation}
Here \( \rho _{S} \) is the nucleon scalar density, \( <\bar{q}q(vac.)> \)
denotes the vacuum expectation value of the condensate and \( \Sigma _{N} \)
is the nucleon sigma term:\begin{equation}
\label{Eq.3.1}
\Sigma _{N}=<N|\left[ Q_{5},\dot{Q}_{5}\right] |N>=2m_{q}\int d\vec{x}<N|\bar{q}q(\vec{x})-\bar{q}q(vac.)|N>.
\end{equation}
 Using the Gellmann, Oakes and Renner relation: \begin{equation}
\label{Eq.4}
f_{\pi }^{2}m_{\pi }^{2}=-2m_{q}<\bar{q}q(vac.)>,
\end{equation}
 the in-medium condensate expression (\ref{Eq.3}) can also be written
in the form:\begin{equation}
\label{Eq.5}
<\bar{q}q(\rho )>=<\bar{q}q(vac.)>+\frac{\Sigma _{N}\rho _{S}}{2m_{q}}.
\end{equation}
 The result above follows from the Feynman-Hellman theorem which relates
the condensate to the thermodynamical grand potential per unit volume
\( \omega =\epsilon -\mu \rho \,  \) through: \begin{equation}
\label{Eq.5.1}
<\bar{q}q(\rho )>={1\over 2}\, \left( {\partial \omega \over \partial m_{q}}\right) _{\mu },
\end{equation}
 where the derivative has to be taken at constant baryonic chemical
potential \( \mu  \) (which controls the density \( \rho  \)). As
an illustration in a free Fermi gas, one has after substracting the
vacuum energy: \begin{equation}
\label{Eq.5.2}
\omega =4\int \frac{d\vec{p}}{(2\pi )^{3}}(E_{p}\, -\, \mu )\, \Theta (\mu -E_{p})
\end{equation}
 with \( E_{p}=\sqrt{p^{2}\, +\, M^{2}} \). The medium contribution
to the condensate is obtained as: \begin{equation}
\label{XXX}
<\bar{q}q(\rho )>-<\bar{q}q(vac.)>={1\over 2}\, \left( {\partial \omega \over \partial M}\right) _{\mu }\, {\partial M\over \partial m_{q}}=\rho _{S}\, {\Sigma _{N}\over 2m_{q}},
\end{equation}
 where the scalar density is defined by:\begin{equation}
\label{RHOS}
\rho _{S}=4\int \frac{d\vec{p}}{(2\pi )^{3}}\, {M\over E_{p}}\, \Theta (\mu -E_{p}),
\end{equation}
 while the ordinary density is:\begin{equation}
\label{RHO}
\rho =4\int \frac{d\vec{p}}{(2\pi )^{3}}\, \Theta (\mu -E_{p})=\frac{2}{3\pi ^{2}}p_{F}^{3}
\end{equation}
with \( p_{F}^{2}=\mu ^{2}-M^{2} \). 

Note that in Eq.(\ref{XXX}) the contribution of the delta function
from the derivative of the Heaviside function vanishes. It will not
be the case for the susceptibility. Starting from Eq.(\ref{Eq.5})
one gets: \begin{equation}
\label{chi_s_rho}
\chi _{S}(\rho )=\frac{\partial }{\partial m_{q}}<\bar{q}q(vac.)>+\rho _{S}\frac{\partial }{\partial m_{q}}\left( \frac{\Sigma _{N}}{2m_{q}}\right) +\frac{\Sigma _{N}}{2m_{q}}\frac{\partial \rho _{S}}{\partial m_{q}},
\end{equation}
which contains three contributions:

\begin{enumerate}
\item The derivative of \( <\bar{q}q(vac.> \), which is the vacuum susceptibility
\( \chi _{S}(vac.) \). Its evaluation would require a non perturbative
QCD model, which is outside the scope of this work. So we focus on
the difference \( \chi _{S}(\rho )-\chi _{S}(vac.). \)
\item The derivative of \( \Sigma _{N}/2m_{q} \) , which is in fact the
nucleon scalar susceptibility%
\footnote{This quantity has not the same dimension as \( \chi _{S} \) due to
the normalisation volume of infinite nuclear matter. 
} \( \chi _{S}^{N}. \) This follows from the relation between the
sigma term and the condensate: \begin{equation}
\label{Eq.5.3}
\Sigma _{N}=2m_{q}\int d\vec{x}<N|\bar{q}q(\vec{x})-\bar{q}q(vac.)|N>.
\end{equation}
 Thus: \begin{equation}
\label{Eq.7}
\chi _{S}^{N}=\frac{\partial }{\partial m_{q}}\int d\vec{x}<N|\bar{q}q(\vec{x})-\bar{q}q(vac.)|N>=\frac{\partial }{\partial m_{q}}\left( \frac{\Sigma _{N}}{2m_{q}}\right) .
\end{equation}
Therefore this second term, which writes \( \rho _{S}\chi _{S}^{N} \)
, can be interpreted as the individual nucleon contribution to the
nuclear susceptibility.
\item The derivative of \( \rho _{S} \) which gives the third term, noted
\( \chi ^{nuclear}_{S} \): \begin{equation}
\label{chi_nuclear}
\chi ^{nuclear}_{S}=\frac{\Sigma _{N}}{2m_{q}}\frac{\partial \rho _{S}}{\partial m_{q}}.
\end{equation}
We shall see that it represents the effect of the nuclear excitations
by contrast to the second term which is due to the nucleon excitations. 
\end{enumerate}
In summary we are going to study separately the two terms of the quantity

\begin{equation}
\label{Eq.6}
\chi _{S}(\rho )-\chi _{S}(vac.)=\rho _{S}\chi _{S}^{N}+\chi ^{nuclear}_{S}.
\end{equation}

\section{Nuclear excitation contribution\label{section-3}}

We first examine the term \( \chi ^{nuclear}_{S} \) in Eq.(\ref{Eq.6}).
We use the free Fermi gas, for which the scalar nucleon density is
given in Eq.(\ref{RHOS}). Taking its derivative with respect to the
quark mass at constant chemical potential we get two terms, the second
one arising from the derivative of the Heaviside function: \begin{equation}
\label{Eq.7.1}
{\partial \rho _{S}\over \partial m_{q}}={\Sigma _{N}\over m_{q}}\, 4\int \frac{d\vec{p}}{(2\pi )^{3}}\, \left[ {p^{2}\over E_{p}^{3}}\, \Theta (E_{p}\, -\, \mu )\, -\, {M^{2}\over E_{p}^{2}}\, \delta (E_{p}\, -\, \mu )\, \right] .
\end{equation}
The first term in Eq.(\ref{Eq.7.1}) represents the polarization through
nucleon-antinucleon (\( \bar{N}N \)) excitations of a relativistic
Fermi gaz submitted to a scalar perturbation (see Appendix). As it
vanishes in the non relativistic limit we shall neglect this \( \bar{N}N \)
contribution.

From the second term in Eq.(\ref{Eq.7.1}) we obtain:\begin{equation}
\label{Eq.7.2}
\chi ^{nuclear}_{S}=\frac{\Sigma ^{2}_{N}}{2m_{q}^{2}}\left( -\frac{2}{\pi ^{2}}\frac{p_{F}M^{2}}{\mu }\right) .
\end{equation}
In the non relativistic limit (\( \mu \simeq M \)) the parenthesis
reduces to \( (-2Mp_{F}/\pi ^{2}) \) which is actually the particle-hole
polarization propagator \( \Pi _{ph}(q) \) of the non relativistic
free Fermi gaz taken in the static situation, \emph{i.e.}, for \( q_{0}=0 \),
and taking the limit of vanishing three momentum \( (\vec{q}\rightarrow 0) \).
We have then: \begin{equation}
\label{Eq.18}
\chi ^{nuclear}_{S}\equiv \frac{\Sigma _{N}}{2m_{q}}\frac{\partial \rho _{S}}{\partial m_{q}}\simeq \frac{\Sigma ^{2}_{N}}{2m_{q}^{2}}\Pi _{ph}(0,\vec{0})\simeq -\frac{2Mp_{F}}{\pi ^{2}}\frac{\Sigma ^{2}_{N}}{2m_{q}^{2}}.
\end{equation}
The presence of \( \Pi _{ph} \) indicates that the origin of this
term lies in the nuclear excitations.

\subsection{Comparison with \protect\( \chi ^{nuclear}_{S}\protect \) obtained
in the sigma model }

\begin{figure}
{\centering \resizebox*{0.6\textwidth}{!}{\includegraphics{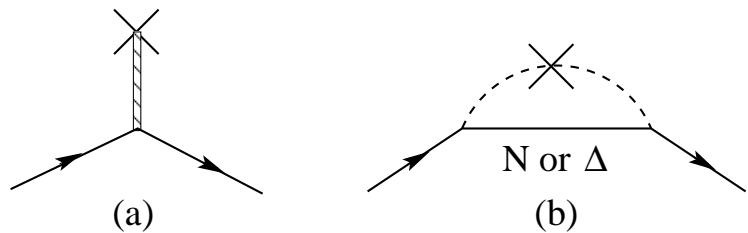}} \par}

\caption{\label{Fig_nuc_sig_term}Nucleon sigma term in the \protect\( \sigma \protect \)
model}
\end{figure}

In the sigma model the scalar susceptibility is related to the \( \sigma  \)
propagator\cite{CE02}. In the medium this propagator is modified
by the particle-hole insertions, which gave:\begin{equation}
\label{Eq.19}
\chi ^{nuclear}_{S}=2\frac{<\bar{q}q(vac.)>^{2}}{f_{\pi }^{2}}\frac{g_{\sigma }^{2}}{m_{\sigma }^{4}}\Pi _{S}(0,\vec{0})
\end{equation}
 where \( \Pi _{S} \) is the full scalar particle-hole polarisation
propagator and \( g_{\sigma } \) is the sigma nucleon coupling constant.
In order to link the two expressions of \( \chi ^{nuclear}_{S} \)
we first need to evaluate the nucleon sigma term \( \Sigma _{N} \)
within the linear sigma model. It is built of two pieces. The first
one corresponds to the exchange of a sigma between the condensate
and the nucleon as illustrated on Figure \ref{Fig_nuc_sig_term}a.
The second one is the contribution of the pion cloud of the nucleon
which comes into play as two pions exchange, as shown in Figure \ref{Fig_nuc_sig_term}b.
The result is\begin{equation}
\label{Eq.20}
\Sigma _{N}=\Sigma _{N}^{\sigma }+\Sigma _{N}^{\pi }=f_{\pi }m_{\pi }^{2}\frac{g_{S}}{m_{\sigma }^{2}}+\frac{m_{\pi }^{2}}{2}\int d\vec{x}<N|\phi ^{2}(\vec{x})|N>.
\end{equation}

\begin{figure}
{\centering \resizebox*{0.6\textwidth}{!}{\includegraphics{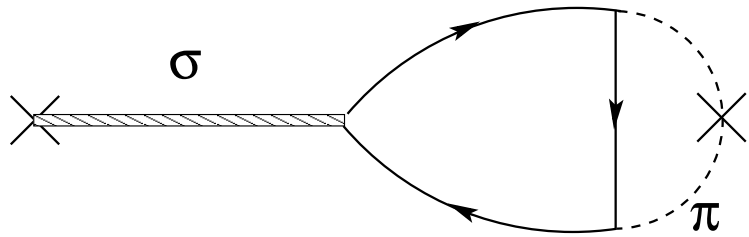}} \par}

\caption{\label{Fig_interf}\.{I}nterference between \protect\( \sigma \protect \)
and \protect\( 2\pi \protect \) exchange. }
\end{figure}

In the context of our previous work\cite{CE02}, the pionic contribution
\( \Sigma _{N}^{\pi } \) did not appear naturally and thus was ignored.
To make the comparison meaningful we should then retain only the sigma
exchange part \( \Sigma _{N}^{\sigma } \) and insert it in our expression
(\ref{Eq.18}) which leads to:\begin{equation}
\label{Eq.21}
\chi _{S}^{nuclear}(\sigma )=\frac{m_{\pi }^{4}f_{\pi }^{2}}{2m_{q}^{2}}\frac{g_{\sigma }^{2}}{m_{\sigma }^{4}}\Pi _{ph}(0,\vec{0})=2\frac{<\bar{q}q(vac.)>^{2}}{f_{\pi }^{2}}\frac{g_{\sigma }^{2}}{m_{\sigma }^{4}}\Pi _{ph}(0,\vec{0}).
\end{equation}
 This is essentially the result of the sigma model calculation (see
Eq.\ref{Eq.19}), provided we replace the full \( \Pi _{S} \) by
the free Fermi gas expression \( \Pi _{ph} \). The present derivation
is more satisfactory in the sense that it does not rely on the sigma
model to derive the coupling between the quark density fluctuations
and the nucleon ones. In particular it incorporates other couplings
than through sigma exchange, such as the two pion exchange term \( \Sigma _{N}^{\pi } \)
of \( \Sigma _{N} \). Moreover the interferences between the various
components of \( \Sigma _{N} \) are automatically incorporated in
the crossed terms of \( \Sigma _{N}^{2} \) . One of these interferences,
that is the \( \Sigma _{N}^{\sigma }\Sigma _{N}^{\pi } \) term, is
illustrated on Figure \ref{Fig_interf}.

On the other hand talking also about the limitations of the present
work, it applies to a free Fermi gas (its generalization is in progress),
while our previous approach did not restrict to this situation. In
the dressing of the sigma line by \( ph \) states, the full \( ph \)
propagator entered. The latter is to some extent constrained by nuclear
phenomenology\cite{CE02} and the argument goes as follows. At normal
nuclear density there is no distinction between the scalar and ordinary
density operators. Now, at ordinary density, the \( ph \) progator
\( \Pi _{\rho } \) is the response of the system to a perturbation
which couples to the nucleon density. In other words it is the nuclear
compressibility, with the relation:\begin{equation}
\label{Eq.21.1}
\Pi _{\rho }=-\frac{9\rho }{K},
\end{equation}
where the nuclear incompressibility \( K \) is related to the energy
per particle \( E/A \) by\begin{equation}
\label{Eq.21.2}
K=\frac{d}{d\rho }\left( \rho ^{2}\frac{d}{d\rho }\frac{E}{A}\right) .
\end{equation}
 Its experimental value at the saturation density \( \rho _{0} \)
is in the range \( 200\div 300\, MeV \) \cite{Blaizot}. This value
is compatible with the free Fermi gas value computed at the same density.
Even though this agreement may result from a cancellation between
several many-body influences, such as the effective mass and residual
interaction effects, it justifies to some extent the use of the free
Fermi gas model.

\section{Nucleon scalar susceptibility \protect\( \chi _{S}^{N}\protect \)\label{section-4}}

For a structureless nucleon one has of course \( \chi _{S}^{N}=0 \)
but we do not make such a restriction. In general \( \chi _{S}^{N}\neq 0 \)
because the true nucleon responds to a scalar perturbation by adjusting
its internal structure. One can estimate this response using models,
for instance the MIT bag model, but it could also be extracted from
lattice calculations when the latter are done at realistic quark mass
values.

\subsection{Valence quark contribution to \protect\( \chi _{S}^{N}\protect \)}

The scalar susceptibility of a free nucleon has been introduced by
Guichon\cite{GSRT96} in another context, concerning a pure nuclear
physics problem, that is the question of saturation of nuclear matter.
In his quark-meson coupling model the saturation follows from the
response of the nucleon to a scalar field. In the bag model the total
scalar charge of the bag defined as:\begin{equation}
\label{Eq.23}
Q_{S}=\int d\vec{x}<N|\bar{q}q(\vec{x})|N>,
\end{equation}
 depends on the quark mass. The linear term in the quark mass expansion
of \( Q_{S} \): \begin{equation}
\label{Eq.24}
Q_{S}(m_{q})=Q_{S}(0)+\chi ^{N}_{S}(Bag)m_{q}+\cdots 
\end{equation}
 defines the susceptibility of the bag. It is found to be \( \chi ^{N}_{S}(Bag)\simeq 0.5R\simeq 2.5\, 10^{-3}MeV^{-1} \),
where the numerical value corresponds to a bag radius \( R=0.8fm \).
It turns out that this susceptibility due to the quark structure can
stabilize the scalar nuclear field and provide a mechanism for saturation\cite{G88}.
In our case however its contribution to the nuclear susceptibility
is negligible (see Section \ref{Numerics}) with respect to the one
due to the nuclear excitations. This is not a surprise since the nucleonic
excitations, which control \( \chi ^{N}_{S}(Bag) \), are much higher
than the nuclear ones. What is interesting however is the positive
sign of this term. Its origin is rather obvious: when its mass increases
the quark becomes less relativistic. This tends to increase the scalar
charge of the nucleon as the quark scalar density \( \bar{q}q \)
involves the difference of the large and small components of the quark
spinor. 

Another important point is that \( \chi ^{N}_{S}(Bag) \) only includes
the valence quark contribution and not those from the pion or sigma
clouds. The argument about the energy of the corresponding excitations
may not apply to the pion cloud contribution in view of the small
pion mass. It is therefore interesting to evaluate the corresponding
susceptibility.

\subsection{Pionic contribution to \protect\( \chi _{S}^{N}\protect \)}

The nucleon sigma commutator is largely influenced by the presence
of the pion cloud. If the nucleon remains unexcited after pion emission
and in the heavy baryon limit the corresponding contribution \( \Sigma _{N}^{\pi } \)
is equal to \cite{JTC92}\begin{equation}
\label{Eq.sigmapi}
\Sigma _{N}^{\pi }=\frac{m_{\pi }^{2}}{2}\int d\vec{x}<N|\phi ^{2}(\vec{x})|N>=\frac{3m_{\pi }^{2}}{16\pi ^{2}}\left( \frac{g_{A}}{f_{\pi }}\right) ^{2}\int _{0}^{\infty }dq\frac{q^{4}}{(q^{2}+m_{\pi }^{2})^{2}}F^{2}(q),
\end{equation}
where \( \phi  \) is the pion field and \( F(q) \) the \( \pi NN \)
form factor. With this explicit expression of \( \Sigma _{N}^{\pi } \)
it is straightforward to calculate the derivatives with respect to
\( m_{q} \) (or \( m_{\pi }^{2} \)) involved in the corresponding
susceptibility:

\begin{eqnarray}
\chi _{S}^{N}(\pi ) & = & \frac{d}{dm_{q}}\frac{\Sigma _{N}^{\pi }}{2m_{q}}=\frac{2<\bar{q}q(vac.)>^{2}}{f_{\pi }^{4}}\frac{d}{dm_{\pi }^{2}}\frac{\Sigma _{N}^{\pi }}{m_{\pi }^{2}}\nonumber \\
 & = & -\frac{3<\bar{q}q(vac.)>^{2}}{4\pi ^{2}f_{\pi }^{4}}\left( \frac{g_{A}}{f_{\pi }}\right) ^{2}\int _{0}^{\infty }dq\frac{q^{4}}{(q^{2}+m_{\pi }^{2})^{3}}F^{2}(q).\label{Eq.pionic.1} 
\end{eqnarray}
For a monopole form factor \( F(q)=\Lambda ^{2}/(\Lambda ^{2}+q^{2}) \)
we find:\begin{equation}
\label{Eq.pionic01}
\chi _{S}^{N}(\pi )=-\frac{9m_{\pi }^{3}}{64\pi }\frac{<\bar{q}q(vac.)>^{2}}{f_{\pi }^{4}}\left( \frac{g_{A}}{f_{\pi }}\right) ^{2}\left( \frac{\Lambda }{\Lambda +m_{\pi }}\right) ^{4}=-4\, 10^{-2}MeV^{-1}
\end{equation}
where the numerical value corresponds to \( \Lambda =5m_{\pi } \).
This value is about 15 times larger than the susceptibility due to
the quark bag structure estimated in the previous section.

A rough estimate of the pionic contribution can be obtained if the
form factor \( F(q) \) is omitted in Eq.(\ref{Eq.pionic.1}), which
corresponds to the limit \( \Lambda \rightarrow \infty . \) In this
case \( \chi _{S}^{N}(\pi ) \) can be written in terms of the leading
non analytical term of the sigma term \( \Sigma _{N}^{LNA} \) according
to:\begin{equation}
\label{Eq.pionic02}
\chi _{S}^{N}(\pi )=\frac{2<\bar{q}q(vac.)>^{2}}{f_{\pi }^{4}}\frac{d}{dm_{\pi }^{2}}\frac{\Sigma _{N}^{LNA}}{m_{\pi }^{2}}=\frac{<\bar{q}q(vac.)>^{2}}{f_{\pi }^{4}m_{\pi }^{4}}\Sigma _{N}^{LNA},
\end{equation}
 where \begin{equation}
\label{Eq.pionic03}
\Sigma _{N}^{LNA}=-\frac{9}{64\pi }\left( \frac{g_{A}}{f_{\pi }}\right) ^{2}m_{\pi }^{3}\simeq -23MeV.
\end{equation}
 This approximation leads to \( \chi _{S}^{N}(\pi )\simeq -8.6\, 10^{-2}MeV^{-1} \),
that is about the double of the value obtained with the form factor. 

There is also a contribution to the nucleon sigma term due to intermediate
\( \pi \Delta  \) states (see Figure \ref{Fig_nuc_sig_term} b).
In the narrow width approximation it reads: \begin{equation}
\label{Eq.sigmapidelta}
\Sigma _{N}^{\pi \Delta }=\frac{3m_{\pi }^{2}}{16\pi ^{2}}\left( \frac{g_{A}}{f_{\pi }}\right) ^{2}\, {4\over 9}\, \left( {g_{\pi N\Delta }\over g_{\pi NN}}\right) ^{2}\, \int _{0}^{\infty }dq\, \left( \frac{q^{4}}{2\omega _{q}^{2}(\omega _{q}+\Delta _{q})^{2}}\, +\, \frac{q^{4}}{2\omega _{q}^{3}(\omega _{q}+\Delta _{q})}\right) \, F^{2}(q)
\end{equation}
 with \( \omega _{q}=\sqrt{q^{2}+m_{\pi }^{2}} \) and \( \Delta _{q}=M_{\Delta }-M_{N}+q^{2}/2M_{\Delta } \).
The corresponding contribution to the susceptibility, \( \chi _{S}^{N}(\pi \Delta ) \),
is obtained from the derivative of the above expression with respect
to \( m_{\pi }^{2} \). The presence of the large energy denominator
\( M_{\Delta }-M_{N} \) makes it less sensitive to the pion mass.
Numerically with the ratio \( g_{\pi N\Delta }/g_{\pi NN}=\sqrt{72}/5 \),
we find \( \chi _{S}^{N}(\pi \Delta )\simeq -1.4\, 10^{-2}MeV^{-1}. \)
The overall susceptibility of a free nucleon due to the pion cloud
is thus about \( -5.4\, 10^{-2}MeV^{-1}. \)

Our conclusion on the nucleon scalar polarisability is that, as the
electric one, it is dominated by the pion cloud. Within a factor of
two it can be expressed in a model independent way in terms of the
leading non analytic part of the nucleon sigma term. In this context
it is legitimate to wonder why the pionic susceptibility \( \chi _{S}^{N}(\pi ) \)
which is so dominant in this problem does not also dominate the nuclear
saturation problem where instead it is totally ignored. The answer
lies in the chiral properties of the scalar field responsible for
the nuclear attraction that we have studied in \cite{CEG02}. We have
stressed that this field has to be chiral invariant rather than \( \sigma  \),
the chiral partner of the pion. It couples derivatively to the pion.
Therefore, in the long wave length and static limit, the pion cloud
is weakly coupled to this nuclear scalar field.

\subsection{Interpretation of the pion cloud contribution in the sigma model.}

\begin{figure}
{\centering \resizebox*{0.6\textwidth}{!}{\includegraphics{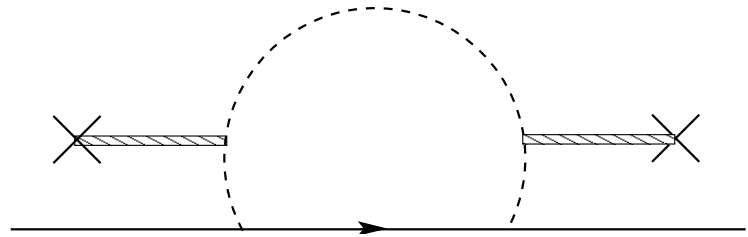}} \par}

\caption{\label{Fig_nuc_susc}Pionic contribution to the nucleon susceptibility.}
\end{figure}

It is interesting to look at the nuclear susceptibility \( \rho _{S}\chi _{S}^{N}(\pi ) \)
in the framework of the sigma model. The sigma, which transmits the
quark fluctuations, is dressed not only by the particle-hole excitations
but also by the two pions excitations, as shown in Figure \ref{Fig_nuc_susc}.
Since we are interested in the modification of the susceptibility
with respect to the its vacuum value, at least one of the pions in
this graph has to be a nuclear one. So it is dressed by particle-hole
insertions. To lowest order in the density the contribution of this
graph to the nuclear susceptibility is:\begin{equation}
\label{Eq.cloud.1}
\chi _{S}^{nuclear,2\pi }=3\frac{2<\bar{q}q(vac.)>^{2}}{f_{\pi }^{2}m_{\sigma }^{4}}\frac{m_{\sigma }^{4}}{4f_{\pi }^{2}}\left( \frac{g_{A}}{f_{\pi }}\right) ^{2}\int \frac{dq^{0}d\vec{q}}{(2\pi )^{4}}\frac{\vec{q}^{2}}{(q^{2}-m_{\pi }^{2})^{3}}\Pi _{L}(q^{0},\vec{q})F^{2}(q).
\end{equation}
Here the subscript \( L \) in \( \Pi _{L} \) indicates the spin
longitudinal character. In the static limit the domain where \( \Pi (q^{0},\vec{q}) \)
has a non vanishing value is pushed to zero energy. In this case the
pions do not carry energy so the integral over \( q^{0} \) which
then involves only \( \Pi (q^{0},\vec{q}) \) reduces to:\begin{eqnarray}
\int \frac{dq_{0}}{2\pi }\Pi _{L}(q_{0},\vec{q}) & = & \int \frac{dq_{0}}{2\pi }\int d\omega \left( -\frac{2\omega }{\pi }\right) \frac{Im\Pi _{L}(\omega \, \vec{q})}{q_{0}^{2}-\omega ^{2}+vi\eta }\nonumber \\
 & = & \int d\omega \left( -\frac{1}{\pi }Im\Pi _{L}(\omega \, \vec{q})\right) =\rho \int d\omega R_{L}(\omega \, \vec{q})=\rho S_{L}(q),\label{Eq-inut1} 
\end{eqnarray}
where \( R_{L}(\omega \, \vec{q})=-Im\Pi _{L}(\omega \, \vec{q})/\pi \rho  \)
is the nuclear longitudinal spin-isospin response and \( S_{L} \)
its integral over energy. For a free Fermi gas one has:\begin{equation}
\label{Eq-inut2}
S_{L}(q)=\Theta (q-2p_{F})+\Theta (2p_{F}-q)\left[ \frac{3}{2}\frac{q}{2p_{F}}-\frac{1}{2}\left( \frac{q}{2p_{F}}\right) ^{3}\right] 
\end{equation}
Ignoring the Pauli blocking effect the quantity \( S_{L}(q) \) reduces
to unity, which leads to: \begin{equation}
\label{Eq.cloud.2}
\chi _{S}^{nuclear,2\pi }=\frac{<\bar{q}q(vac.)>^{2}}{2f_{\pi }^{4}}\rho \frac{d}{dm_{\pi }^{2}}\left( \frac{\Sigma _{N}^{\pi }}{m_{\pi }^{2}}\right) =\rho \chi _{S}^{N}(\pi ).
\end{equation}
 This is exactly the nucleonic polarisability arising from the pion
cloud multiplied by the density. In fact the mere comparison of the
many-body graph of Figure \ref{Fig_sig_2pi} with the one of Figure
\ref{Fig_nuc_susc} which represents the free nucleon susceptibility
leads to this conclusion. To leading order the evaluation of the nuclear
QCD scalar susceptibility arising from the \( 2\pi  \) intermediate
states does not need any calculation as it is simply related to the
nucleon one, if Pauli blocking is ignored.

\section{Numerical estimates\label{Numerics}}

We have now all ingredients to proceed to the numerical evaluation
of the in-medium modification of the scalar susceptibility. To fix
the idea we shall use \( \Sigma _{N}\simeq 45\, MeV \) and \( K\simeq 230MeV^{-1} \).
At normal density the contribution of the nuclear excitations to the
susceptibility is then: \[
\chi ^{Nuclear}_{S}(\rho _{0})=-8.2\, 10^{5}\, MeV^{2}.\]
Turning now to the nucleonic participation to the scalar susceptibility
we have to take into account the Pauli blocking which reduces the
pionic cloud contribution from \( \pi N \) intermediate states by
about \( 25\% \) at normal density, bringing the in-medium nucleon
susceptibility to the effective value \( \tilde{\chi }_{S}^{N}\simeq -4.9\, 10^{-2}MeV^{-1}. \)
We multiply it by the nuclear density which gives \( \rho _{0}\tilde{\chi }_{S}^{N}\simeq -6.8\, 10^{4}MeV^{2}. \)
This number is smaller than the nuclear excitation contribution. Altogether
the scalar susceptibility of nuclear matter at normal density is:\begin{equation}
\label{numeric.3}
\chi _{S}(\rho _{0})=\chi _{S}(vac.)-8.9\, 10^{5}MeV^{2}.
\end{equation}
It is interesting to give a scale to compare this nuclear modification
of the scalar susceptibility. The susceptibility of the vacuum \( \chi _{S}(vac.) \)
is not known but due to the large mass of the scalar meson, it is
certainely much smaller than the pseudoscalar susceptibility \( \chi _{PS}(vac) \).
The latter is actually dominated by the Goldstone boson, i.e. the
pion, which allows its evaluation. Chanfray and Ericson give the following
expression\cite{CE02}: \begin{equation}
\label{numeric.4}
\chi _{PS}(vac)=-\frac{2<\bar{q}q(vac.)>^{2}}{f_{\pi }^{2}m_{\pi }^{2}}\simeq -1.3\, 10^{6}MeV^{2}.
\end{equation}
From this we infer that: 

\begin{itemize}
\item \( \chi _{S(vac.)} \) can reasonnably be neglected on the right hand
side of Eq.(\ref{numeric.3}).
\item at \( \rho =\rho _{0} \) the nuclear scalar susceptibility is comparable
to the vacuum pseudoscalar susceptibility \( \chi _{PS}(vac) \).
\end{itemize}
Moreover Chanfray and Ericson have shown\cite{CE02} that \( \chi _{PS}(\rho ) \)
follows the density evolution of the quark condensate, \emph{i.e.},
at normal density it has decreased by \( 35\% \), which brings it
to \( \chi _{PS}(\rho _{0})\simeq -9\, 10^{5}MeV^{2} \). This is
nearly equal to the value \( \chi _{S}(\rho _{0})\simeq -8.9\, 10^{5}MeV^{2} \)
which we get when we neglect \( \chi _{S(vac.)} \) in Eq.(\ref{numeric.3}).
This implies that the scalar and pseudoscalar susceptibilities, which
are so different in the vacuum, become nearly equal at normal nuclear
density, a feature normally expected only near the chiral phase transition.
As our evaluation uses the value of the free nucleon sigma term, this
convergence of \( \chi _{S}(\rho _{0}) \) and \( \chi _{PS}(\rho _{0)} \)
toward a common value must be taken with a grain of salt due to the
possible medium renormalisation: \( \Sigma _{N}\rightarrow \tilde{\Sigma }_{N}(\rho _{0}) \),
which we have not taken into account in this work. However this manifestation
of the restoration of chiral symmetry is so spectacular that we do
not expect it to be totally destroyed by these renormalisation effects.
A systematic investigation of this problem, as well as the extension
of this study at larger densities, deserves further work. 

In summary we have found that the two QCD susceptibilities, namely
the scalar and pseudoscalar ones, undergo a strong modification in
the nuclear medium in such a way they become close to each other already
at normal nuclear matter density. It is a spectacular consequence
of partial chiral restoration which may have consequences in processes
involving higher densities.

\section*{Appendix\label{Appendix_A}}

Here we want to interpret the first term in Eq.(\ref{Eq.7.1}) in
terms of nucleon antinucleon excitations. For this we can consider
a relativistic free fermi gas described by a hamiltonian \( H_{0} \)
and add a perturbation%
\footnote{In the general case one should start with a space dependent perturbation
and let it tend to a constant at the end. For the \( N\bar{N} \)
contribution this step is not necessary.
}\begin{equation}
\label{A1}
\lambda W=\lambda \int d\vec{r}\bar{\psi }\psi =\lambda \int d\vec{r}\, \rho _{S}(\vec{r})
\end{equation}
which changes the nucleon mass by the amount \( \delta M=\lambda  \).
If we note \( |\lambda > \) the ground state of the system in presence
of the perturbation then, by the Feynamn Hellman theorem, we have\begin{equation}
\label{A2}
\frac{\partial }{\partial \lambda }\left. <\lambda |\int d\vec{r}\, \rho _{s}(\vec{r})|\lambda >\right| _{\lambda =0}=V\frac{\partial }{\partial \lambda }\left. <\lambda |\rho _{S}(0)|\lambda >\right| _{\lambda =0}=2<\lambda =0|W\frac{1}{E_{0}-H_{0}}W|\lambda =0>,
\end{equation}
where we assume \( <\lambda |\lambda >=1 \) for simplicity and \( V \)
is the volume of the gas. To simplify we note \begin{equation}
\label{A3}
\frac{\partial }{\partial \lambda }\left. <\lambda |\rho _{S}(0)|\lambda >\right| _{\lambda =0}=\frac{\partial \rho _{S}}{\partial \lambda }.
\end{equation}
From the cannonical field expansion, the part of \( W \) which produces
the \( N\bar{N} \) intermediate states is \begin{eqnarray}
W_{N\bar{N}} & = & \int d\vec{p}\frac{1}{2E_{p}}[\bar{u}(\vec{p})v(-\vec{p})b^{\dagger }(\vec{p})d^{\dagger }(-\vec{p})]+h.c.\nonumber \\
 & = & -\int d\vec{p}\frac{\vec{\sigma }.\vec{p}}{E_{p}}[b^{\dagger }(\vec{p})d^{\dagger }(-\vec{p})]+h.c.\label{A4} 
\end{eqnarray}
 where \( b,\, d \) are respectively the destruction operators of
the nucleon and antinucleon. So the \( N\bar{N} \) contribution to
the R.H.S\.{ } of Eq.(\ref{A2}) is:

\begin{eqnarray}
\left. \frac{\partial \rho _{S}}{\partial \lambda }\right| _{N\bar{N}} & = & \frac{2}{V}<\lambda =0|W_{N\bar{N}}\frac{1}{E_{0}-H_{0}}W_{N\bar{N}}|\lambda =0>\nonumber \\
 & = & \frac{2}{V}<\lambda =0|\int d\vec{p}\frac{\vec{\sigma }.\vec{p}}{E_{p}}b(\vec{p})d(-\vec{p})\frac{1}{E_{0}-H_{0}}\int d\vec{p}'\frac{\vec{\sigma }.\vec{p}'}{E_{p}'}d^{\dagger }(-\vec{p}')b^{\dagger }(\vec{p}')|\lambda =0>\nonumber \\
 & = & \frac{2}{V}\int d\vec{p}\, \frac{p^{2}}{E_{p}^{2}}\left( \frac{1}{-2E_{p}}\right) <\rho |b(\vec{p})b^{\dagger }(\vec{p})|\rho >\nonumber \\
 & = & \frac{1}{V}\int d\vec{p}\, \frac{p^{2}}{E_{p}^{3}}<\rho |b^{\dagger }(\vec{p})b(\vec{p})|\rho >-C_{\infty }\nonumber \\
 & = & \frac{4}{(2\pi )^{3}}\int ^{p_{F}}_{0}d\vec{p}\, \frac{p^{2}}{E_{p}^{3}}-C_{\infty },\label{A5} 
\end{eqnarray}
where we have used\begin{equation}
\label{A6}
<\lambda =0|b^{\dagger }(\vec{p})b(\vec{p})|\lambda =0>=4\delta (\vec{0})\theta (p_{F}-p)=\frac{4V}{(2\pi )^{3}}\theta (p_{F}-p).
\end{equation}
The infinite term \( C_{\infty } \) is independent of the density
so it drops out when we substract the vacuum contribution. Since the
perturbation (\ref{A1}) is equivalent to a change \( \delta M=\lambda  \)
of the nucleon mass, we can write\begin{equation}
\label{A7}
\left. \frac{\partial \rho _{S}}{\partial m_{q}}\right| _{N\bar{N}}=\frac{\partial M}{\partial m_{q}}\left. \frac{\partial \rho _{S}}{\partial \lambda }\right| _{N\bar{N}}=\frac{\Sigma _{N}}{m_{q}}\frac{4}{(2\pi )^{3}}\int ^{p_{F}}_{0}d\vec{p}\, \frac{p^{2}}{E_{p}^{3}},
\end{equation}
 which is the first term in Eq.(\ref{Eq.7.1}). In other terms the
derivative of the scalar density at fixed density is entirely due
to the \( N\bar{N} \) excitations.

Note that this does not allow us to say that the term\begin{equation}
\label{A8}
\frac{\Sigma _{N}}{2m_{q}}\left. \frac{\partial \rho _{S}}{\partial m_{q}}\right| _{N\bar{N}}
\end{equation}
in Eq.(\ref{chi_nuclear}) is the nuclear susceptibility due to \( N\bar{N} \)
excitations. To reach such a conclusion we should start with a perturbation
of the form\begin{equation}
\label{A9}
m_{q}\int d\vec{r}[\bar{u}u(\vec{r})+\bar{d}d(\vec{r})]=m_{q}W,
\end{equation}
that is the true mass term of QCD. Defining the nuclear susceptibility
(per unit volume) as\begin{equation}
\label{A10}
\chi ^{nuclear}_{S}=\frac{1}{2V}\left. \frac{\partial }{\partial m_{q}}<m_{q}|\int d\vec{r}[\bar{u}u(\vec{r})+\bar{d}d(\vec{r})]|m_{q}>\right| _{m_{q}=0}
\end{equation}
where \( |m_{q}> \) denotes the nuclear ground state in presence
of the quark mass term (\ref{A9}) then the Feynman-Hellman theorem
gives\begin{equation}
\label{A11}
\chi ^{nuclear}_{S}=\left( \frac{1}{2V}\right) 2<m_{q}=0|W\frac{1}{E-H_{0}}W|m_{q}=0>.
\end{equation}
To compute this quantity using the Fermi gas approximation we need
to know the matrix element \( <N|W|N> \) for the \( ph \) excitations
and \( <N\bar{N}|W|vac.> \) for the \( N\bar{N} \) excitations.
There is no problem with the first one since we know it from the nucleon
sigma term \( \Sigma _{N} \) . On the other hand the second one is
essentially unknown. We can quantify this problem by introducing the
\( NN \) and \( N\bar{N} \) scalar form factors in the standard
way:\begin{eqnarray}
m_{q}<N(p')|\bar{u}u(0)+\bar{d}d(0)|N(p)> & = & S^{NN}[(p-p')^{2}]\bar{u}(p')u(p),\nonumber \\
m_{q}<N(p')\bar{N}(p)|\bar{u}u(0)+\bar{d}d(0)|vac.> & = & S^{N\bar{N}}[(p+p')^{2}]\bar{u}(p')v(p)\label{A12} 
\end{eqnarray}
From the definition of the nucleon sigma term:\begin{equation}
\label{A13}
\Sigma _{N}=\frac{1}{<N(0)|N(0)>}<N(0)|\int d\vec{r}\, m_{q}[\bar{u}u(\vec{r})+\bar{d}d(\vec{r})]|N(0)>,
\end{equation}
 we get \( S^{NN}(0)=\Sigma _{N} \) and by the crossing rule \( S^{N\bar{N}} \)
and \( S^{NN} \) are the same function. So we can write \begin{equation}
\label{A14}
<N(p')\bar{N}(p)|\bar{u}u(0)+\bar{d}d(0)|vac.>=\frac{\Sigma _{N}}{m_{q}}g[(p+p')^{2}]\bar{u}(p')u(p)
\end{equation}
 where we have defined \( g(x)=S^{NN}(x)/S^{NN}(0) \). A straightforward
calculation then leads to the following expression%
\footnote{After substraction of the vacuum contribution
} for the \( N\bar{N} \) contribution to the nuclear susceptibility:\begin{equation}
\label{A15}
\chi ^{nuclear}_{S}(N\bar{N})=\frac{\Sigma _{N}^{2}}{2m_{q}^{2}}\frac{4}{(2\pi )^{3}}\int ^{p_{F}}_{0}d\vec{p}\left| g(4E_{p}^{2})\right| ^{2}\frac{p^{2}}{E_{p}}\approx \left| g(4M^{2})\right| ^{2}\frac{\Sigma _{N}}{2m_{q}}\left. \frac{\partial \rho _{S}}{\partial m_{q}}\right| _{N\bar{N}}
\end{equation}
which differs from (\ref{A8}) by the factor \( \left| g(4M^{2})\right| ^{2} \)\( . \)
This factor is probably very small because the transition \( vacuum\, \rightarrow \, N\bar{N} \)
through the one body quark operator \( \bar{u}u+\bar{d}d \) is suppressed
as compared to the elastic transition \( N\rightarrow N. \)

\newpage

\end{document}